 \documentclass[nofootinbib,aps,prl,preprint,longbibliography]{revtex4-1}
\usepackage{amsmath}
\usepackage{bbold}
\usepackage{graphicx}
\usepackage[pdfencoding=auto]{hyperref}
\usepackage{bookmark}
\usepackage{lineno}
\newcommand{\etal}{\textit{et al}.}

\begin{document}

\author{Leonie Spitz$^{1,2}$}
\email{leonie.spitz@psi.ch}
\author{Takuya Nomoto$^{3}$}
\author{Shunsuke Kitou$^{1}$}
\author{Hironori Nakao$^{4}$}
\author{Akiko Kikkawa$^{1}$}
\author{Sonia Francoual$^{5}$}
\author{Yasujiro Taguchi$^{1}$}
\author{Ryotaro Arita$^{1,3}$}
\author{Yoshinori Tokura$^{1,3,6}$}
\author{Taka-hisa Arima$^{1,7}$}
\author{Max Hirschberger$^{1,3}$}

\affiliation{$^1$RIKEN Center for Emergent Matter Science (CEMS), Wako, Saitama 351-0198, Japan}
\affiliation{$^2$Current address: Paul Scherrer Institute, 5232 Villigen PSI, Switzerland}
\affiliation{$^3$Department of Applied Physics and Quantum-Phase Electronics Center (QPEC), The University of Tokyo, Bunkyo-ku, Tokyo 113-8656, Japan}
\affiliation{$^4$Institute of Materials Structure Science, High Energy Accelerator Research Organization, Tsukuba, Ibaraki 305-0801, Japan}
\affiliation{$^5$Deutsches Elektronen-Synchrotron DESY, Notkestraße 85, 22607 Hamburg, Germany}
\affiliation{$^6$Tokyo College, The University of Tokyo, Bunkyo-ku, Tokyo 113-8656, Japan}
\affiliation{$^7$Department of Advanced Materials Science, University of Tokyo, Kashiwa, Chiba 277-8561, Japan}

\email{hirschberger@ap.t.u-tokyo.ac.jp}
\date{\today}

\title{Entropy-assisted, long-period stacking of honeycomb layers in an AlB$_2$-type silicide}
\maketitle

\begin{large}
 \begin{center}
    Abstract
\end{center}   
\end{large}

Configurational entropy can impact crystallization processes, tipping the scales between structures of nearly equal internal energy. Using alloyed single crystals of Gd$_2$PdSi$_3$ in the AlB$_2$-type structure, we explore the formation of complex layer sequences made from alternating, two-dimensional triangular and honeycomb slabs. A four-period and an eight-period stacking sequence are found to be very close in internal energy, the latter being favored by entropy associated with covering the full configuration space of interlayer bonds. Possible consequences of polytype formation on magnetism in Gd$_2$PdSi$_3$ are discussed.

\section{Introduction}
The study of configurational entropy in crystallization processes is experiencing a revival with research into the stabilization of high-entropy alloys and oxides \cite{Yeh2004, George2019, Oses2020}. A more established playground for the exploration of entropic effects during crystallization is the chemistry of polytypes, which characterize stacking sequences of geometrically equivalent layers with different repeat period \cite{Verma1966,Trigunayat1989,Durovic1992}. The impact of vibrational entropy\cite{Jagodzinski1954,Jagodzinski1995} and entropy gains due to phase separation during crystallization \cite{Kim2005} has been previously scrutinized for stacking processes of layers. As for configurational entropy in layer stacking, its importance has been largely neglected, especially in the case of rigid layers: researchers thought that configurational entropy, which should scale roughly proportional to the sample thickness, is overwhelmed by energetic contributions proportional to the sample volume \cite{Wahab1980,Trigunayat1989}. 

Our target material is $R_2$PdSi$_3$ ($R$: rare earth element), which derives from the AlB$_2$ hexagonal close-packed structure type, and which is attracting attention for its magnetic properties as a host of complex skyrmion spin-textures \cite{Kurumaji2019,Hirschberger2020a, Hirschberger2020b} [Fig. \ref{fig:fig1}]. This structure is defined by alternating rare earth and Pd/Si slabs, which form perfectly planar triangular and honeycomb sheets, respectively. We control long-period polytype formation by alloying Gd$_2$Pd(Si$_{1-x}$Ge$_x$)$_3$ (GPSG) and use high-resolution synchrotron x-ray scattering to determine the stacking sequence. The eight-layer stack of $R_2$PdSi$_3$, shown in Fig. \ref{fig:fig1}, is the longest-period polytype out of all AlB$_2$-derived silicides \cite{Nentwich2020a, Nentwich2020b,Tang2011}. Its thermodynamic origins remain mysterious, especially because Coulomb interactions between distant layers are strongly suppressed by charge screening in metals. Instead, configurational entropy is found to tip the balance between short- and long-period polytypes of Gd$_2$PdSi$_3$. We observe discommensuration within a single layer, which limits the characteristic size of structurally ordered blocks, and allows the entropic contribution to scale in proportion to the sample volume.

\section{Experimental Methods}
Large single crystals of the solid solution Gd$_2$Pd(Si$_{1-x}$Ge$_x$)$_3$ are grown in a floating zone furnace, ranging from $x = 0$ to $x = 0.5$. At the highest $x$ studied here, the viscosity of the melt zone is drastically reduced by the presence of Germanium. Figure \ref{fig:fig2}(a) shows a representative rod, with single-crystalline domains of dimensions around $5$ centimeters. The power supplied to a halogen lamp in the furnace, shown in Fig. \ref{fig:fig2}(b), serves as a proxy for the temperature of the molten zone. For our alloyed crystals, the melting temperature drops rapidly as the content of Germanium increases. 

For $x = 0$, $0.1$, and $0.3$, we report single crystal diffraction data measured at beamline P09 of PETRA III at DESY~\cite{Strempfer2013}. Single crystalline samples with polished surfaces are mounted in a displex cryostat with base temperature $T_\mathrm{base}=5\,$K. Data are collected in multi-bunch mode, using a pyrolytic graphite (PG-006) analyzer plate to suppress fluorescent background. Measurements are carried out in the vertical scattering geometry with incident photons linearly polarized in the plane perpendicular to the scattering plane ($\sigma$-incident photons). We focus on measurements in the $I_{\sigma-\sigma'}$ channel, selecting the diffracted photons polarized in the plane perpendicular to the scattering plane, as is expected for charge scattering. We found that the structural scattering observed here at x-ray energy $E = 7.934\,$keV is nearly independent of $E$~\cite{SI}.

For $x = 0.4$, room-temperature experiments are carried out using a four-circle diffractometer with $18\,$kW rotation-anode x-ray source at the High Energy Accelerator Research Organization (KEK). The Mo K$\alpha$ radiation, $E = 17.5\,$keV, is monochromatized by PG-002. Lattice constants and $(HKL)$ indices in reciprocal space are defined with respect to the 1H unit cell in Fig. \ref{fig:fig1}(b), i.e., with respect to the elemental building block of the polytypes discussed in this work. Some previous work on Gd$_2$PdSi$_3$ defines lattice constants $a$, $c$ with respect to the original AlB$_2$ cell of Fig. 1(a), where the in-plane lattice constant is $a(\mathrm{AlB_2}) = a(\mathrm{1H})/2$.

\section{Results}
\begin{figure}[htb]
  \centering
  \includegraphics[clip, width=1.\linewidth]{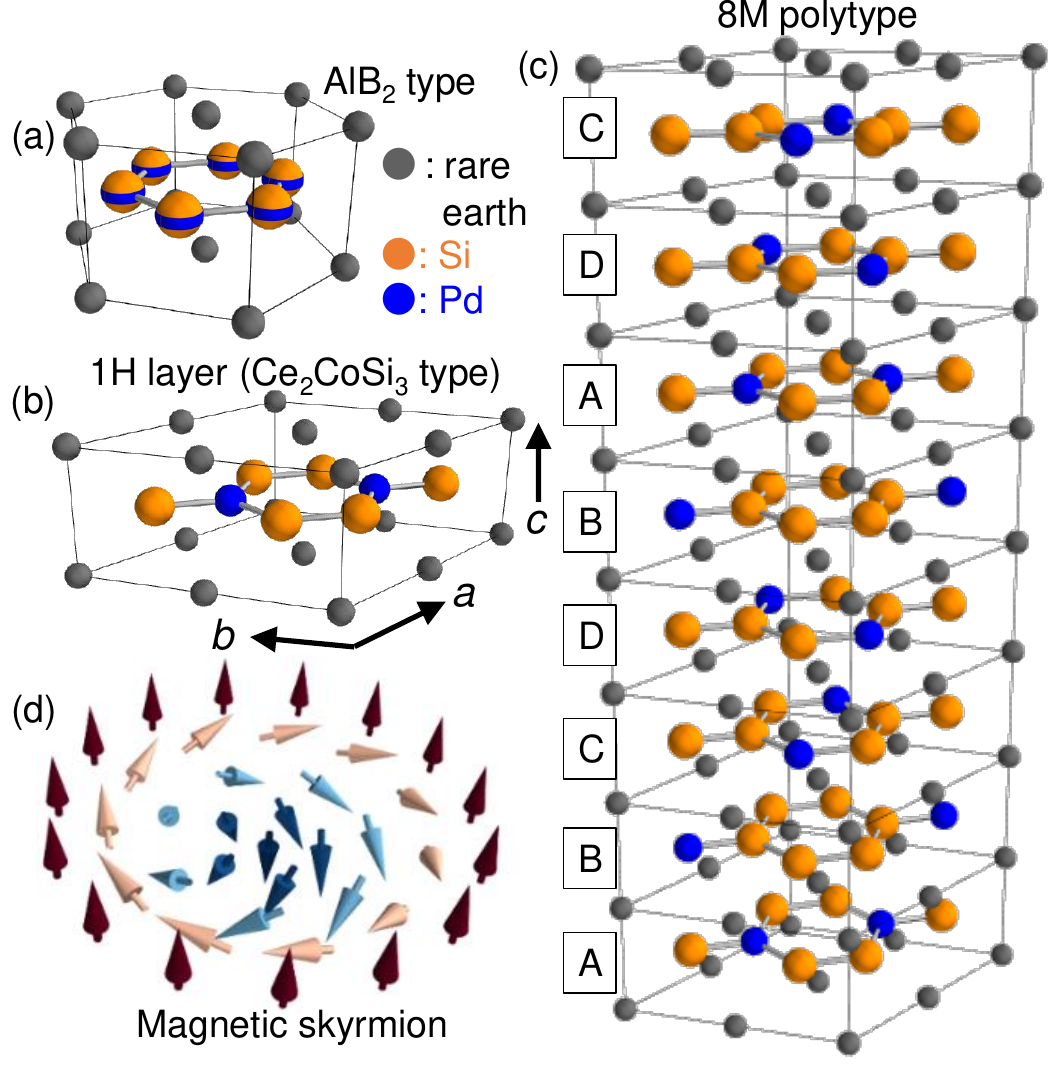}
  \caption{Crystal structure and stacking order of Gd$_2$PdSi$_3$. (a) Hypothetical AlB$_2$-type structure with mixed site occupancy on the honeycomb layer. (b) Gd$_2$PdSi$_3$ (1H) unit cell with Pd/Si ordering. (c) Centrosymmetric 8M superstructure of Gd$_2$PdSi$_3$, with eight type-1H layers~\cite{Tang2011}. (d) Magnetic skyrmion, a spin texture realized in this centrosymmetric material family.}
\label{fig:fig1}
\end{figure}

\begin{figure}[b]
  \centering
  \includegraphics[clip, trim=0.2cm 0.3cm 0.2cm 0.2cm, width=0.8\linewidth]{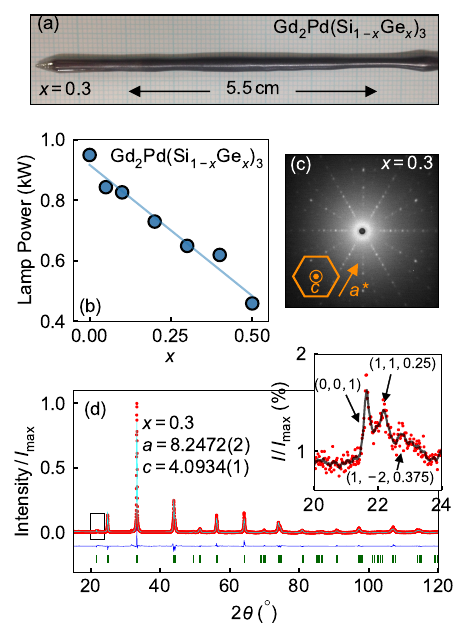}
  \caption{Characterization of Gd$_2$Pd(Si$_{1-x}$Ge$_x$)$_3$ single crystals. (a) Optical image of Gd$_2$Pd(Si$_{0.7}$Ge$_{0.3}$)$_3$ grown by the floating zone technique. (b) Lamp power in floating zone growth as a function of Germanium content $x$. The line is a guide to the eye. (c) Laue x-ray image of a single crystal with $x=0.3$, with x-ray beam along the $c$-direction. (d) Powder x-ray diffraction (red) at $x = 0.3$, with Rietveld refinement (cyan), residual of the fit (blue), and expected reflections (green vertical lines) . Inset: low-angle range, where we identify weak reflections related to layer-stacking along the $c$-direction. A smoothed trace of the data is shown in black.}
\label{fig:fig2}
\end{figure}

\begin{figure}[htb]
  \centering
  \includegraphics[clip, trim=0.2cm 0.3cm 0.1cm 0.2cm, width=0.8\linewidth]{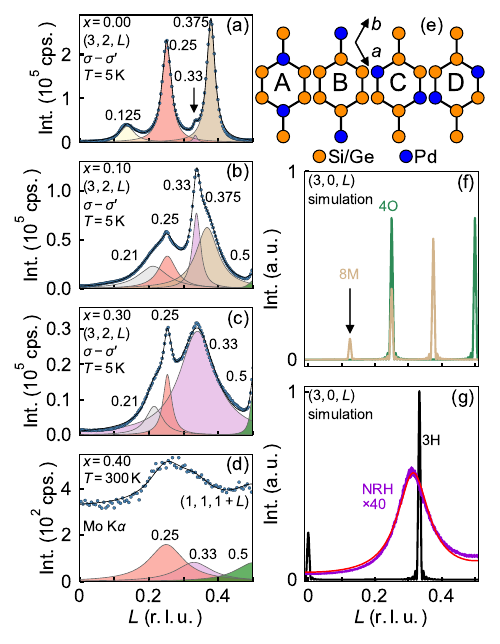}
  \caption{Layer stacking along the $c$-direction in Gd$_2$Pd(Si$_{1-x}$Ge$_{x}$)$_3$ from synchrotron x-ray diffraction. (a-d) Reciprocal space L-scan around ($3,2,0$) at photon energy $E = 7934\,$eV. Several maxima of intensity are identified and fitted with Lorentzian line profiles. (e)  Four building blocks (A-D) of the polytypes 4O (ABCD), 8M (ABCDBADC), 3H (ABC), as well as a nearly-random, short-range correlated sequence (NRH). Only the honeycomb layer is shown. (f,g) Numerical simulation of scattering intensity, assuming four types of structures assembled from elements in (e); see text for discussion.}
\label{fig:fig3}
\end{figure}

Figure \ref{fig:fig1} builds up the long-period stacking sequence of Gd$_2$PdSi$_3$, starting from the AlB$_2$-type with random occupation on the honeycomb layer. This hypothetical Gd(Pd$_{1/4}$Si$_{3/4}$)$_2$ is unfavorable due to effective Coulomb repulsion between Pd-Pd~\cite{Tang2011,Nentwich2020a, Nentwich2020b}. Instead, Pd-Si ordering emerges in the honeycomb plane, where Pd atoms occupy positions of maximal mutual distance [Fig. \ref{fig:fig1}(b)]. Given the geometry of the honeycomb lattice with six sites for Pd/Si, four types of layers A, B, C, D are possible [Fig. \ref{fig:fig1}, Fig. \ref{fig:fig3}(e)]. The uniform stacking of any one of these, termed 1H in conventional polytype notation, has not yet been observed. Tang \etal{} reported the monoclinic eight-layer stack of Fig. \ref{fig:fig1}(d), described here as 8M. The shown ABCDBADC sequence is only one of six coexisting symmetry equivalent domains~\cite{Tang2011,Frontzek2010}. Space group symbols for 8M and other stacking patterns are given in the Supplementary Information~\cite{SI}.

Initially, as-grown single crystals are thoroughly characterized via in-house laboratory x-ray techniques. We show a Laue pattern, representative of the high quality of the present single crystals, in Fig. \ref{fig:fig2}(c). The x-ray beam is aligned along the $c$-direction, revealing a six-fold pattern characteristic for the averaged hexagonal symmetry. Finally, we present x-ray diffraction data of powders made from crushed single-crystalline pieces in Fig. \ref{fig:fig2}(d), with a full Rietveld refinement~\cite{SI}. The major peaks are indexed with respect to the 1H structure in Fig. 1(b), where a series of weaker reflections (inset) is associated with polytype formation, or stacking along the $c$-axis.

Precise information about the stacking sequences can be gathered from diffuse x-ray diffraction observed in experiments on single crystals (Fig. \ref{fig:fig3}). A series of broad reflections appears along the $(3,2,L)$ line of reciprocal space in the parent compound Gd$_2$PdSi$_3$: specifically $L=n/8$ with $n = 1, 2, 3$, as well as weak intensity at $L=1/3$ that is consistent with prior work~\cite{FrontzekPhD2009}. We use a Lorentzian line-shape to extract the center position and peak-width of these reflections, summarized in Table 1. Scattering intensity around $L=1/3$ drastically increases when introducing disorder to the Pd/Si honeycomb sublattice via Germanium alloying, while the $1/8$ and $3/8$ reflections become much harder to discern. Moreover, intensity develops at $L = 0.5$ when $x$ is increased.

\begin{table}[hbt!]
\centering
\setlength{\belowcaptionskip}{10pt}
\caption{Lorentzian widths of peaks related to short-range order along the $(3, 2, L)$ line, in reciprocal lattice units (r.l.u.). The corresponding raw data and fits are shown in Fig. \ref{fig:fig3}. NRH marks the simulation of a nearly random, yet short-range correlated stack of hexagonal symmetry, which is also shown in Fig. \ref{fig:fig3}(g). In line-scans along the $L$-direction, the linewidths of fundamental Bragg reflections of the 1H structure for $x=0, 0.1, 0.3$ are observed to be $\sigma= 0.005\,$r.l.u., and $\sigma = 0.008$ for $x=0.4$ (Supplementary Information).} 
\begin{tabular}{ c | c c c c c }
 x & 1/8 & 1/4 & 1/3 & 3/8 & 1/2  \\
\hline 
 0.0 & 0.05 & 0.03 & 0.013 & 0.03 & - \\  
 0.1 & - & 0.06 & 0.03 & 0.08 & 0.03 \\  
 0.3 & - & 0.03 & 0.13 & - & 0.02\\  
 0.4 & - & 0.12 & 0.12 & - & 0.12 \\
NRH & - & - & 0.12 & - & -      
\end{tabular}
\label{tab:peakform}
\end{table}

We use a numerical approach to calculate the diffracted intensity expected for several types of stacking patterns. X-ray intensities are simulated by Fourier transforming a virtual lattice (dimensions $200\,a\times 200\,a\times 120 \,c$), to obtain the scattering intensity via the standard expression $I(q)\propto \left|A(q)\right|^2$, with $A(q) = \sum_{j\in\text{u.c.}}f_j\exp(i\mathbf{q}\cdot\mathbf{r}_j)$, where $j$ counts over all atoms in the virtual lattice. The complex number $f_j$ is the atomic form factor, which depends on $|\mathbf{q}|$ and the atomic species, and whose imaginary part grows large close to atomic absorption edges. The contribution of atomic displacements due to thermal vibration is ignored. Note that Gadolinium atoms do not contribute to the diffuse scattering in our model, as -- contrary to Pd/Si in the honeycomb layer -- their position does not change from layer to layer.

Every simulated stacking sequence is built from four types of layers A-D, shown in Fig. \ref{fig:fig3}(e), also taking into account the presence of symmetry-equivalent domains. The polytypes 4O (ABCD, orthorhombic) and 8M (e.g. ABCDBADC, monoclinic) are two simple sequences incorporating A-D in equal proportion, where direct Pd-Pd and Pd-Si-Pd bonds are avoided. The conspicuous weakness of scattering intensity at $L = 0$, $0.5$ in the parent compound $x = 0$ [Fig. \ref{fig:fig3}(a)] severely constrains our modeling~\cite{Tang2011}. Only the 8M structure has near-zero intensity at these positions. Meanwhile, $x = 0.3$ and $x=0.4$ have finite volume fractions of 4O stacking, judging from the intensity ratios $I(L = 0.5)/ I(L = 0.25)$ in Fig. \ref{fig:fig3}(c, d). 

Intensity around $L = 0.33$ appears in two distinct models considered in our calculations. Figure \ref{fig:fig3}(g) displays calculated intensities for a 3H (ABC) stack, which is a chiral, inversion-breaking arrangement of layers -- in contrast to the centrosymmetric 4O and 8M patterns in Fig. \ref{fig:fig3}(f). The significant intensity at $L=0$ predicted for 3H is not observed in any of our experiments, suggesting that the chiral structure is at best a minority phase in our crystals. 
Instead, we propose that a random sequence with short-range repulsion between Pd atoms can generate strong intensity around $L=1/3$, labeled as NRH (nearly random, hexagonal). NRH is generated numerically by initializing a random sequence of layers, and subsequently eliminating blocks such as A-A, A-B-A, and equivalent. The simulation is shown in Fig. \ref{fig:fig3}(g) as a purple curve, with the red line indicating a Lorentzian fit to the simulated data. Table \ref{tab:peakform} includes this fit result. Significantly, NRH results in a ratio $r_0 = I_{L=0}/I_{max}<6\,\%$, while 3H has $r_0=25\,\%$. $I_{max}$ is the most intense Bragg peak measured.

In summary, the data suggest that long-period 8M stacking gives way to a roughly even mixture of two types of sequences at $x = 0.3$: the nearly random sequence NRH, and 4O. At $x = 0.4$, where the melting temperature of the single crystal is significantly reduced [Fig. \ref{fig:fig1}(b)], the 4O stacking dominates. 
Moving further to $x=0.5$, we detect diffuse scattering using an area detector, with broad streaks of intensity centered around $L = 0.33$, consistent with NRH (Supplementary Information).



\section{Discussion}
Alloying experiments are suitable to shed light on the hitherto elusive origin of the 8M stacking sequence in stoichiometric Gd$_2$PdSi$_3$, which is the longest-period polytype in AlB$_2$-derived silicides~\cite{Nentwich2020a,Nentwich2020b}. All data are consistent with the notion that the fundamental building blocks of Gd$_2$Pd(Si$_{1-x}$Ge$_x$)$_3$, at least up to $x = 0.5$, are the 1H layers A-D depicted in Fig. \ref{fig:fig3}(e). We compare several ordered stacking sequences, or polytypes built from 1H, based on energetic and entropic considerations. Firstly, ferroic stacks 1H and two-layer sequences (e.g. AB) result in Pd-Pd or Pd-Si-Pd bonds along the $c$-direction. Ab-initio calculations confirm that Pd-Pd bonds are highly unfavorable in energy, both within a single honeycomb layer~\cite{Tang2011} and between subsequent layers~\cite{SI}.

\begin{figure}[htb]
  \centering
	\includegraphics[clip, trim=5.5cm 4.2cm 8.1cm 5cm, width=0.8\linewidth]{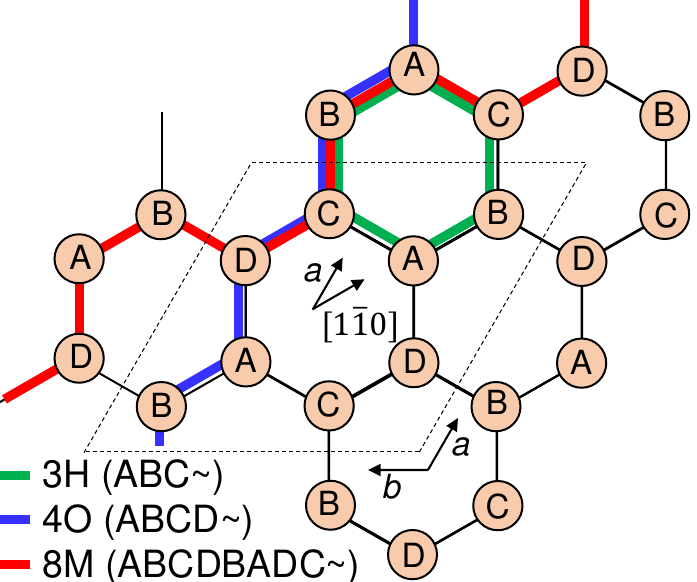}
  \caption{Stacking patterns of four types of honeycomb layers (A-D) projected onto one-dimensional graphs in a 2D honeycomb plane. The letters on each honeycomb site indicate the position of Pd in the respective layers. We illustrate 3H, 4O, and 8M stacking patterns as one-dimensional graphs in green, blue, and red, respectively. Each bond connects between two subsequent layers: for example, the green 3H graph connects vertically between stacked ABC$\sim$ layers.}
\label{fig:fig4}
\end{figure}

Second, the chiral sequence 3H with three-layer period necessarily causes internal strain, as  illustrated in Fig. \ref{fig:fig4}. The four types of layers in Fig. \ref{fig:fig3}(e) have Pd atoms on different positions in the honeycomb network. Hence, we may illustrate a stacking sequence as a linear graph on a projected honeycomb network. Every vertex is labeled by the layer A-D for which it is occupied by Palladium. Bonds between vertices connect nearest neighbors in the stacking sequence. In this graph, the chiral 3H stack corresponds to a loop, e.g. the green line in Fig. \ref{fig:fig4}. Some vertices of the honeycomb network are never reached by the 3H graph. These positions correspond to infinite vertical chains of Si atoms in the crystal, which surround hexagons partially occupied by Palladium atoms. The spatial separation of Pd and Si drives internal strain.

Thirdly, we may compare the longer-period polytypes 4O and 8M by their graphs on the projected honeycomb plane in Fig. \ref{fig:fig4}. Both correspond to infinite lines, indicating breaking of 1H's hexagonal symmetry. We predict the formation of at least three types of domains. The resulting weak orthorhombic / monoclinic crystal distortion is directed along $a$ in the case of 4O, and along $a^*$ in the case of 8M. As for lifting the degeneracy between 4O and 8M, long-range Coulomb interactions over distances of four layers are strongly suppressed by electrostatic screening, and the difference in ground state energies is small on the level of density functional theory~\cite{SI}. Hence, we consider the contribution of entropy $S$ to the free energy $F = U-TS$, where $U$ and $T$ are internal energy and temperature, respectively. It is noticeable in Fig. \ref{fig:fig4} that the armchair graph (8M) makes use of a larger number of different bonds. Meanwhile the zigzag graph (4O) only makes use of two bonds, so that we naively expect configurational entropy to favor 8M. 

The problem of decorating a lattice with a given set of atomic species was previously studied by Burdett et al.~\cite{Burdett1985a,Burdett1985b} It was realized that the configuration of lowest energy can be determined by finding the shortest disparate loop (length $n$) on the lattice, i.e. the shortest loop that distinguishes the two decorations. In the present case, $n\ge 6$ when comparing 4O and 8M polytypes, and in fact loops with $n = 8$ are expected to make the dominant contribution to the Coulombic energy difference. This large value of $n$ reinforces our position that $\Delta U$ between 4O and 8M is so small that extensive entropy can become significant.

We attempt a more quantitative estimate of the entropy gain from formation of 8M. In the x-ray scattering data, broad diffuse scattering appears in the $(H,K,0)$ plane\cite{SI}. Such intensity indicates dislocations within a single honeycomb layer. The correlation length within a honeycomb layer is around $20$ lattice constants of the AlB$_2$ parent structure, i.e. $\sim 8\,$nm, allowing us to estimate the entropy gain $\Delta S$ of 8M over 4O. At a melting temperature around $1690\pm 40\,^\circ$C, we estimate $T \Delta S \approx 2.3\,$meV / $V_{uc}$, where we use the volume of an 8M unit cell [Fig. \ref{fig:fig1}(d)] as a reference. 

The effect of alloying is two-fold: On the one hand, introduction of Ge reduces the melting temperature of our crystals, necessarily suppressing the entropic effect. On the other hand, random occupation of Ge and Si in the honeycomb layers is expected to weaken interlayer correlations, and to disfavor the monoclinic distortion; potentially opening the door towards realizing the chiral polytype 3H. Indeed, our $x = 0.1$ sample may show a minority phase of 3H, although the much broader reflections around $L = 1/3$ observed at $x = 0.3$, $0.4$, and $0.5$ are attributed to a nearly random stacking sequence (NRH)~\cite{SI}.

\section{Conclusion and Outlook}
The present study reports the stabilization of an eight-layer polytype 8M in Gd$_2$PdSi$_3$ over competing stacking patterns via configurational entropy. We demonstrate the controllability of polytype formation in Gd$_2$Pd(Si$_{1-x}$Ge$_x$)$_3$ (GPSG) by chemical alloying, and illustrate the role of a slight structural distortion in ruling out the competing chiral polytype 3H. As compared to the chemistry of high entropy alloys, where random occupation of several atomic species on a lattice is promoted by configurational entropy, the present polytype problem approximates the question of stacking ordered layers of four flavors (A, B, C, D). GPSG hence offers a simplified, quasi one-dimensional materials platform to explore the impact of configurational entropy on structural chemistry.

Physical properties of materials rarely change from one polytype to another \cite{Verma1966,Luo2015}, although the fundamental symmetries of the lattice may differ. In the present case, one may speculate about the effect of polytype formation on magnetic interactions in Gd$_2$PdSi$_3$, underpinning the formation of complex spin textures. The spin Hamiltonian of this material was recently studied by Paddison \textit{et al.} using inelastic neutron scattering on a powder sample~\cite{Paddison2022}. These authors suggest that further-neighbor exchange interactions play an important role in stabilizing the skyrmion spin texture~\cite{Paddison2022}. In a competing scenario, skyrmions in centrosymmetric solids are stabilized by magnetic exchange anisotropy, which is allowed in Gd$_2$PdSi$_3$ only if the triangular lattice plane of Gd-atoms is not a global mirror plane~\cite{Hirschberger2021}. Polytype formation is important in both cases: In the former, structural polytype formation modulates the exchange interaction along the $c$-direction, potentially affecting the stability range of the skyrmion phase~\cite{Paddison2022}. In the latter case, mirror symmetry breaking in the 4O and 8M structures is the key to unlocking the power of anisotropic exchange in Gd$_2$PdSi$_3$~\cite{Hayami2021a,Hayami2021b}. Control of polytype formation thus adds a new twist to the field of skyrmion research in centrosymmetric solids, setting Gd$_2$PdSi$_3$ apart from related compounds such as hexagonal Gd$_3$Ru$_4$Al$_{12}$~\cite{Hirschberger2019} as well as tetragonal GdRu$_2$Si$_2$~\cite{Khanh2020,Khanh2022} and EuAl$_4$~\cite{Takagi2022}. 

\section{Supporting Information} The supporting information describes growth and characterization of single crystals, a measurement of the melting point of Gd$_2$PdSi$_3$, additional x-ray scattering data and experimental details, and ab-initio calculations of the electronic structure. Diffuse scattering data of Gd$_2$Pd(Si$_0.5$Ge$0.5$)$_3$, obtained at BL-02B1 of SPring-8 using a PILATUS area detector, is provided in two-dimensional data sets. 

\section{Acknowledgments} We acknowledge discussions with S. Gao. L.S. was funded by the German Academic Exchange Service (DAAD) via a PROMOS scholarship awarded by the German Federal Ministry of Education and Research (BMBF). M.H. benefited from support by JSPS KAKENHI Grant Nos. JP21K13877 and JP22H04463, as well as the Iketani Foundation. This work was partially supported by JST CREST Grant Nos. JPMJCR1874 and JPMJCR20T1, and JST-PREST Grant No. JPMJPR20L7 (Japan). Synchrotron radiation experiments were performed at SPring-8 with the approval of the Japan Synchrotron Radiation Research Institute (JASRI) (Proposal No. 2021B1261), and at PETRA-III at Deutsches Elektronen-Synchrotron (DESY), a Research Center of the Helmholtz Association (HGF), under proposal I-20190781 EC.

\bibliography{GPSG_Structure}

\end{document}